\documentclass{elsart}
\usepackage{amssymb}
\def\1{{_{1}}}
\def\2{{_{2}}}
\def\3{{_{3}}}
\def\5{{_{5}}}
\def\7{{_{7}}}

\begin{document}
%%%%%%%%%%%%%%%%%%%%%%%%%%%%%%%%%%%%%%%%%%

\runauthor{Jizba and Arimitsu}
\begin{frontmatter}
\title{Towards information theory for $q$-nonextensive statistics
without $q$-deformed distributions
% A murder A murder A murder A murder A murder A murder A murder A murder A
%murder
%\thanksref{X}
}
\author[Praha]{Petr Jizba\thanksref{*}} and
\author[Tsukuba]{Toshihico Arimitsu}
%\author[Tsukuba]{Catullus}
%\author[Baiae]{Publius Maro Vergilius}
%\author[Paestum]{Unknown author}
\address[Praha]{FNSPI, Czech Technical University, B\v{r}ehov\'{a} 7,
115 19 Praha 1, Czech Republic}
\address[Tsukuba]{Institute of Physics, Tsukuba University, Ibaraki
305-8571, Japan}
%\address[Baiae]{The White House, Baiae}
%\address[Rome]{Senate House, Rome}
\thanks[*]{Corresponding author: Tel.: +420-224-35-8320;  Fax: +420-222-32-0861\\{$\mbox{\hspace{1.5mm}}$\em ~~E-mail addresses:}
p.jizba@fjfi.cvut.cz (P.J.); arimitsu@cm.ph.tsukuba.ac.jp (T.A.)}
%\thanks[**]{{\footnotesize{{\em E-mail addresses:} p.jizba@fjfi.cvut.cz
%(P.J.);
%arimitsu@cm.ph.tsukuba.ac.jp (T.A.).}} }

\begin{abstract}
\begin{minipage}{15.5cm}
{\hspace{4mm}} In this paper we extend our recent results [P.Jizba,
T.Arimitsu Physica~{\bf A~340}~(2004)~110] on $q$-nonextensive
statistics with non-Tsallis entropies. In particular, we combine an
axiomatics of R\'{e}nyi with the $q$-deformed version of Khinchin
axioms to obtain the entropy which accounts both for systems with
embedded self-similarity and $q$-nonextensivity. We find that this
entropy can be uniquely solved in terms of a one-parameter family of
information measures. The corresponding entropy maximizer is
expressible via a special function known under the name of the
Lambert W-function. We analyze the corresponding ``high" and
``low-temperature" asymptotics and make some remarks on the
possible applications.\\
\\
{\em PACS:~} 65.40.Gr, 47.53.+n, 05.90.+m     \\[3mm]
{\em Keywords:} Information theory; R{\'e}nyi's information
measure; Tsallis--Havrda--Charv\'{a}t \\entropy; MaxEnt principle;
Lambert W-function
\end{minipage}
\end{abstract}
%\begin{keyword}
%Information theory; R{\'e}nyi's information measure;\\
%Tsallis-Havrda-Charv\'{a}t entropy; MaxEnt principle; Lambert
%W-function \PACS
%\end{keyword}
\end{frontmatter}
\vspace{-7mm}
%
%%%%%%%%%%%%%%%%%%%%%%%%%%%%%%%%%%%%%%%%
\section{Introduction}\label{I}
%%%%%%%%%%%%%%%%%%%%%%%%%%%%%%%%%%%%%%%%
%
\vspace{-4mm}
The idea that Gibbsian statistical thermodynamics and Shannon's
communication theory share the same line of reasoning was
originally introduced by Edwin~Jaynes in his influential 1957
papers~\cite{Jaynes57}. There he proposed the {\em Maximum Entropy
Principle} (MaxEnt) as a general inference procedure with a direct
relevance to statistical physics. A standard frame of statistical
thermodynamics appeared as soon as the notion of entropy was
introduced. In particular, Jaynes's MaxEnt utilized Shannon's
entropy, or better, Shannon's information measure ($= S$), as an
inference functional. The central r\^{o}le of Shannon's entropy as
a tool for inductive inference (i.e., inference where prior
information are given in terms of expectation values) was further
demonstrated in works of Faddeyev~\cite{Fad1}, Shore and
Johnson~\cite{SJ1}, Wallis~\cite{Wal1} and others. Following
Jaynes, one should view the MaxEnt distribution (or maximizer) as
a distribution that is maximally noncommittal with regard to
missing information and that agrees with all what is known about
prior information, but expresses maximum uncertainty with respect
to all other matters~\cite{Jaynes57}.
%*** By identifying the statistical sample
%space with the set of all (coarse-grained) microstates the
%corresponding maximizer yields the Shannon entropy that
%corresponds to the Gibbs entropy of statistical thermodynamics.

With the advancement in information theory it has become clear that
Shannon's entropy is not the only feasible information measure.
Indeed, many modern communication processes, including signals,
images and coding systems, often operate in complex environments
dominated by conditions that do not match the basic tenets of
Shannon's communication theory. For instance, buffer memory (or
storage capacity) of a transmitting channel is often finite, coding
can have a non-trivial cost function, codes might have
variable-length codes, sources and channels may exhibit memory or
losses, etc. Post-Shannon developments of information theory offer
various generalized measures of information to deal with such
situations. Measures of Havrda-Charv\'{a}t~\cite{HaCh},
Sharma-Mittal~\cite{SM1}, R\'{e}nyi's~\cite{Re2} and
Kapur's~\cite{Kap1} can serve as examples.
%The situation gets yet
%more complex when communication systems with quantum channels are
%considered~\cite{}.

If the parallel between information theory and statistical physics
has a deeper reason, as advocated by Jaynes, then one should expect
similar progress also in statistical physics. Indeed, in the past 20
years, physicists have begun to challenge the assumptions of Gibbs's
statistics such as ergodicity or metric transitivity. This happened
when evidence accumulated showing that there are many situations of
practical interest requiring statistics which do not conform with
Gibbs's exponential maximizers. Examples include percolation, cosmic
rays, turbulence, granular matter, clustered volatility, etc.

When trying to generalize Gibbs's entropy, the
information-theoretic parallel with statistical thermodynamics
provides a useful conceptual guide. The natural strategy that fits
this framework is to revisit the axiomatic rules governing
Shannon's entropy and potential extensions translate into a
language of statistical physical. The usual axiomatics of
Khinchin~\cite{Kh1} is prone to several cogent generalizations.
Among those, the additivity of independent mean information is a
natural axiom to attack. In this way, two fundamentally distinct
generalization schemes have been pursued in the literature; one
redefining the statistical mean and another generalizing the
additivity rule. While the first leads to R\'{e}nyi's
entropies~\cite{Re2,Re1} that are nature tool in systems with
embedded self-similarity~\cite{PJ1}, the second scheme yields
various deformed entropies~\cite{Naudts} that play important
r\^{o}le in long-range/time correlated systems.

It is to be expected that a suitable merger of the above
generalizations could provide a new conceptual frame suitable for
a statistical description of systems possessing both
self-similarity and long-range correlations. Such systems are
quite pertinent with examples spanning from the early cosmological
phase transitions to currently much studied quantum phase
transitions (frustrated spin systems, Fermi liquids, etc.). Our
aim was to study  one particular merger, namely merger of
R\'{e}nyi and Tsallis--Havrda--Charv\'{a}t (THC) entropies.

The structure of this paper is the following: In Section~II
axiomatics of R\'{e}nyi and THC entropies are reviewed. In Section
III we formulate a new axiomatics which aims at unifying the
R\'{e}nyi and THC entropies. Such an axiomatics allows for only
one one-parameter family of information measures. Basic properties
of this new class of entropies are discussed in Section IV. The
ensuing maximizer is calculated in Section V. There we show that
MaxEnt distribution is expressible through the Lambert W-function.
We analyze the corresponding ``high" and ``low-temperature"
asymptotics and discuss the corresponding non-trivial structure of
the parameter space. A final discussion is given in Section VI.
\vspace{-7mm}
%
%%%%%%%%%%%%%%%%%%%%%%%%%%%%%%%%%%%%%%%%%%%%%%%%%%%%%%%%%%%%%
\section{R\'{e}nyi's and THC entropies --- axiomatic viewpoint}
\label{IV}
%%%%%%%%%%%%%%%%%%%%%%%%%%%%%%%%%%%%%%%%%%%%%%%%%%%%%%%%%%%%%%
%
\vspace{-4mm}
As already said, RE represents a step towards more realistic
situations encountered in information theory. Since RE's have a
firm operational characterization given in terms of block coding
and hypotheses testing (see, e.g.,~\cite{Csi1}), it can be
directly measured. This is typically happening, e.g., in
communication systems with the buffer overflow problem or in
variable-length coding with an exponential cost constraint.
%R\'{e}nyi parameter $q$ then
%represents the so-called $\beta$-cutoff rates~\cite{Csi1}.
RE's are also indispensable in various branches of physics that
require self-similar sample spaces. Examples being chaotic
dynamical systems or multifractals. RE of order $q$ that is
assigned to a discrete distribution ${\mathcal{P}} = \{ p_1,
\ldots, p_n\} $ is defined as\\[-8mm]
\begin{eqnarray}
{\mathcal{I}}_q({\mathcal{P}}) = \frac{1}{(1 - q)} \ln\left(
\sum_{k=1}^n (p_k)^q  \right)\, , \;\;\;\;\;\;\; q \ > \ 0\, .
\end{eqnarray}
For simplicity's sake we use the base $e$ of natural logarithms.
RE thus defined is then measured in natural units --- nats, rather
than bits.\footnote{To convert, note that $1$ bit $=0.693$ nats.}

%Fully develop turbulence~\cite{}, earthquake analysis~\cite{} and
%generalized dimensions of strange attractors~\cite{} provide
In his original work R\'{e}nyi~\cite{Re1} introduced a
one-parameter family of information measures ($=$RE)  which he
based on axiomatic considerations. His axioms have been further
sharpened by Dar\'{o}tzy~\cite{Dar1} and others~\cite{Oth2}.  It
has been recently shown in Ref.~\cite{PJ1} that RE can be
conveniently characterized by the following set of axioms:
\begin{enumerate}
\item For a given integer $n$ and given ${\mathcal{P}} = \{ p_1,
p_2, \ldots , p_n\}$ ($p_k \geq 0, \sum_k^n p_k =1$),
${\mathcal{I}}({\mathcal{P}})$ is a continuous with respect to all
its arguments.

\item For a given integer $n$, ${\mathcal{I}}(p_1, p_2, \ldots ,
p_n)$ takes its largest value for $p_k = 1/n$ ($k=1,2, \ldots, n$)
with the normalization ${\mathcal{I}}\left( \frac{1}{2},
\frac{1}{2}\right) = \ln 2$.
%${\mathcal{I}}_{\alpha}({\mathcal{P}})$ takes its largest value for
%$p_k = 1/n, (k = 1,2, \ldots, n)$, i.e., the gained information is
%largest when we known least about the original system.

\item For a given $q\in {{\mathbb{R}}}$; $~{\mathcal{I}}(A\cap B)
= {\mathcal{I}}(A) + {\mathcal{I}}(B|A)$ with\\
${\mathcal{I}}(B|A) = \mbox{{\textsl{g}}}^{-1} \left(\sum_k
\varrho_k(q) \mbox{{\textsl{g}}}({\mathcal{I}}(B|A=A_k)) \right)$,
and $\varrho_k(q) = p_k^{q}/\sum_k p_k^{q}$ with $ p_k =
{\mathcal{P}}(A_k)$.

\item $\mbox{{\textsl{g}}}$ is invertible and positive in $[0,
\infty)$.

\item ${\mathcal{I}}(p_1,p_2, \ldots , p_n, 0 ) =
{\mathcal{I}}(p_1,p_2, \ldots , p_n)$, i.e., adding an event of
probability zero (impossible event) we do not gain any new
information.
\end{enumerate}
\noindent These axioms markedly differ from those utilized
in~\cite{Re1,Dar1,Oth2}. Important distinction is the emergence of
the zooming (or escort) distribution $\varrho(q)$ in axiom 3. Note
also that RE of two independent experiments is additive. In fact,
it was proved in Ref.~\cite{Re1} that RE is the most general
information measure compatible with additivity of independent
information and the Kolmogorov system of probability.
%RE found a
%direct interpretation in coding theory with arbitrary cost
%function~\cite{Campbell}. It helps to estimate Lyapunov exponents
%and dimension spectra and can be easier connected with
%experimental parameters than the Kolmogorov-Sinai entropy~\cite{}.
%
%\vspace{-7mm}
%
%%%%%%%%%%%%%%%%%%%%%%%%%%%%%%%%%%%%%%%%%%%%%%%%%%%%%%%%%%%%%%%%%
%\section{THC entropy: entropy of long-distance/time correlated
%systems}\label{V}
%%%%%%%%%%%%%%%%%%%%%%%%%%%%%%%%%%%%%%%%%%%%%%%%%%%%%%%%%%%%%%%%%
%
%\vspace{-4mm}

Among variety of deformed entropies the currently popular one is
the $q$-deformed Shannon's entropy, better known as  THC entropy.
As the classical additivity of independent information is not
valid there,
%a new more exotic physical mechanisms must be sought
%to comply with THC predictions.
one may infer that the typical playground for THC entropy should
be in systems with non-vanishing long-range/time correlations:
e.g., in statistical systems with quantum non-locality or in
various option-price models. In the case of discrete distributions
${\mathcal{P}} = \{ p_1, \ldots, p_n \}$ THC entropy takes the
form:\\[-8mm]
\begin{eqnarray}
{\mathcal{S}}_q({\mathcal{P}}) \ = \ \frac{1}{(1-q)} \left[
\sum_{k=1}^n (p_k)^q  -1\right]\, , \;\;\;\;\;\;\; q \ > \ 0\, .
\end{eqnarray}
\noindent Axiomatic treatment was recently proposed in
Ref.~\cite{Ab2} and it consists of four axioms
\begin{enumerate}
\item For a given integer $n$ and given ${\mathcal{P}} = \{ p_1,
p_2, \ldots , p_n\}$ ($p_k \geq 0, \sum_k^n p_k =1$),
${\mathcal{S}}({\mathcal{P}})$ is a continuous with respect to all
its arguments.

\item For a given integer $n$, ${\mathcal{S}}({\mathcal{P}})$
takes its largest value for $p_k = 1/n$ ($k=1,2, \ldots, n$).
%${\mathcal{I}}_{\alpha}({\mathcal{P}})$ takes its largest value for
%$p_k = 1/n, (k = 1,2, \ldots, n)$, i.e., the gained information is
%largest when we known least about the original system.

\item For a given $q\in {\mathbb{R}}$; ${\mathcal{S}}(A\cap B) =
{\mathcal{S}}(A) + {\mathcal{S}}(B|A) +
(1-q){\mathcal{S}}(A){\mathcal{S}}(B|A)$ with
 ${\mathcal{S}}(B|A) =
\sum_k \varrho_k(q) \ {\mathcal{S}}(B|A=A_k)$.

%For independent events, i.e., ${\mathcal{R}} = {\mathcal{P}}\times
%{\mathcal{Q}}=
%\{p_iq_k \}$: \\
%${\mathcal{I}}_{\alpha}({\mathcal{P}}\times {\mathcal{Q}}) =
%{\mathcal{I}}_{\alpha}({\mathcal{P}}) + {\mathcal{I}}_{\alpha}({\mathcal{Q}})$.
%
%\item In general case there is a continuous invertible function
%$f(x)$ such that: ${\mathcal{I}}_{\alpha}\partial
%f({\mathcal{I}}_{\alpha})/\partial
%{\mathcal{I}}_{\alpha} = g({\mathcal{I}}_{\alpha}) f({\mathcal{I}}_{\alpha})$.\\
%\\
%Here ${\mathcal{I}}_{\alpha}({\mathcal{P}}\cup {\mathcal{Q}}) =
%{\mathcal{I}}_{\alpha}({\mathcal{I}}_{\alpha}({\mathcal{P}}),
%{\mathcal{I}}_{\alpha}(\mathcal{Q}))$. The scaling function $g(x)$ is
%common to both ${\mathcal{I}}_{\alpha}({\mathcal{P}}\cup {\mathcal{Q}}),
%{\mathcal{I}}_{\alpha}({\mathcal{P}})$ and ${\mathcal{I}}_{\alpha}({\mathcal{Q}})$
%.

\item ${\mathcal{S}}(p_1,p_2, \ldots , p_n, 0 ) =
{\mathcal{S}}(p_1,p_2, \ldots , p_n)$.
\end{enumerate}
As said before, one keeps here the linear mean but generalizes the
additivity law. In fact, the additivity law in axiom~3 is nothing
but the Jackson sum
%(or $q$--additivity)
of the $q$ calculus.
\vspace{-7mm}
%
%%%%%%%%%%%%%%%%%%%%%%%%%%%%%%%%%%%%%%%%%%%%%%%%%%%
\section{Axiomatic merger}\label{VII}
%%%%%%%%%%%%%%%%%%%%%%%%%%%%%%%%%%%%%%%%%%%%%%%%%%%
%
\vspace{-4mm}
As a natural axiomatic merger of previous two axiomatics one can
choose:
\begin{enumerate}
\item For a given integer $n$ and given ${\mathcal{P}} = \{ p_1,
p_2, \ldots , p_n\}$ ($p_k \geq 0, \sum_k^n p_k =1$),
${\mathcal{D}}({\mathcal{P}})$ is a continuous with respect to all
its arguments.

\item For a given integer $n$, ${\mathcal{D}}({\mathcal{P}})$
takes its largest value for $p_k = 1/n$ ($k=1,2, \ldots, n$).
%${\mathcal{I}}_{\alpha}({\mathcal{P}})$ takes its largest value for
%$p_k = 1/n, (k = 1,2, \ldots, n)$, i.e., the gained information is
%largest when we known least about the original system.

\item For a given $q\in {\mathbb{R}}$; ${\mathcal{D}}(A\cap B) =
{\mathcal{D}}(A) + {\mathcal{D}}(B|A) +
(1-q){\mathcal{D}}(A){\mathcal{D}}(B|A)$ with ${\mathcal{D}}(B|A)
= f^{-1}\left(\sum_k \varrho_k(q) \
f\left({\mathcal{D}}(B|A=A_k)\right)\right)$.

%For independent events, i.e., ${\mathcal{R}} = {\mathcal{P}}\times
%{\mathcal{Q}}=
%\{p_iq_k \}$: \\
%${\mathcal{I}}_{\alpha}({\mathcal{P}}\times {\mathcal{Q}}) =
%{\mathcal{I}}_{\alpha}({\mathcal{P}}) + {\mathcal{I}}_{\alpha}({\mathcal{Q}})$.
%
%\item In general case there is a continuous invertible function
%$f(x)$ such that: ${\mathcal{I}}_{\alpha}\partial
%f({\mathcal{I}}_{\alpha})/\partial
%{\mathcal{I}}_{\alpha} = g({\mathcal{I}}_{\alpha}) f({\mathcal{I}}_{\alpha})$.\\
%\\
%Here ${\mathcal{I}}_{\alpha}({\mathcal{P}}\cup {\mathcal{Q}}) =
%{\mathcal{I}}_{\alpha}({\mathcal{I}}_{\alpha}({\mathcal{P}}),
%{\mathcal{I}}_{\alpha}(\mathcal{Q}))$. The scaling function $g(x)$ is
%common to both ${\mathcal{I}}_{\alpha}({\mathcal{P}}\cup {\mathcal{Q}}),
%{\mathcal{I}}_{\alpha}({\mathcal{P}})$ and ${\mathcal{I}}_{\alpha}({\mathcal{Q}})$
%.

\item $f$ is invertible and positive in $[0, \infty)$.

\item ${\mathcal{D}}(p_1,p_2, \ldots , p_n, 0 ) =
{\mathcal{D}}(p_1,p_2, \ldots , p_n)$.
\end{enumerate}
In Refs.~\cite{PJ2,PJ3} it has been shown that the above axioms
allow for only one one-parameter class of solutions given by\\[-8mm]
\begin{eqnarray}
{\mathcal{D}}_q({\mathcal{A}}) \ &=& \   \frac{1}{1-q}\ \left(
e^{-(1-q)\sum_k \varrho_k(q) \ln p_k}  -1 \right) \ = \
\frac{1}{1-q} \left( \prod_k (p_k)^{-(1-q)\varrho_k(q)} -1
\right)
%\nonumber
%\\[2mm]
%&=& \ \ln_{\{q\}}\left(
%e^{-\langle\ln{\mathcal{P}}\rangle_q}\right)
\, . \label{VIe}
\end{eqnarray}
Here $\langle \ldots \rangle_q$ is defined with respect to the
distribution $\varrho_k(q)$. We can further recast the relation
(\ref{VIe}) into another, more convenient, form:\\[-8mm]
\begin{eqnarray}
{\mathcal{D}}_q({\mathcal{A}}) \ = \ \frac{1}{1-q}\ \left(
e^{-(1-q)^2 d{\mathcal{I}}_q/dq} \sum_{k=1}^n (p_k)^q -1 \right)\,
. \label{VIf}
\end{eqnarray}
Eqs.(\ref{VIe})--(\ref{VIf}) represent the sought information
measure.
\vspace{-5mm}
%
%%%%%%%%%%%%%%%%%%%%%%%%%%%%%%%%%%%%%%%%%%%%%%%%%%%%%%%%%%
\section{Basic properties of ${\mathcal{D}}_q$}
%%%%%%%%%%%%%%%%%%%%%%%%%%%%%%%%%%%%%%%%%%%%%%%%%%%%%%%%%%
%
\vspace{-4mm}
Before studying the implications of the formulas
(\ref{VIe})--(\ref{VIf}), there is one immediate consequence which
warrants special mention. In particular, from the condition
$d{\mathcal{I}}_q/dq \leq 0$ (see, e.g.,~\cite{Re2}) one has\\[-8mm]
\begin{eqnarray}
{\mathcal{D}}_q(A) \; \left\{
\begin{array}{ll}   \geq  \ {\mathcal{S}}_q(A) \;
 & \mbox{if ~ $q \leq 1$}\\
\leq \ {\mathcal{S}}_q(A) \; & \mbox{if ~ $q \geq 1$}
\end{array} \right. \, ,
\end{eqnarray}
with equality, iff  $q=1$ or $d{\mathcal{I}}_q/dq =0$. This
happens only when ${\mathcal{P}}$ is uniform or trivial, i.e., $\{
1,0, \ldots, 0\}$. By utilizing the known properties of
${\mathcal{I}}_q$ and ${\mathcal{S}}_q$ we have
\begin{eqnarray}
&&0 \ \leq \ {{S}}({\mathcal{P}}) \ \leq \
{\mathcal{I}}_q({\mathcal{P}}) \ \leq \
{\mathcal{S}}_q({\mathcal{P}})\ \leq \
{\mathcal{D}}_q({\mathcal{P}}) \ \leq \ \ln_q n,
\,\,\,\,\,\,\,\,\, \mbox{for} \,\, 0 < q \leq 1\, ,\nonumber \\
&&  0 \ \leq \ {\mathcal{D}}_q({\mathcal{P}}) \ \leq \
{\mathcal{S}}_q({\mathcal{P}}) \ \leq \
{\mathcal{I}}_q({\mathcal{P}})\ \leq \ {{S}}({\mathcal{P}}) \ \leq
\ \ln n, \,\,\,\,\,\,\,\,\,\,\,\, \mbox{for} \,\, q \ \geq \ 1\, .
\label{IIIl}
\end{eqnarray}
This means that by investigating the information measure
${\mathcal{D}}_q$ with the given $q < 1$ we receive more
information than restricting to ${\mathcal{I}}_q$ or
${\mathcal{S}}_q$ only. On the other hand, when $q > 1$ then both
${\mathcal{I}}_q$ and ${\mathcal{S}}_q $ are more informative than
${\mathcal{D}}_q$. In practice one usually requires more than one
$q$ to gain more complete information about a system. In fact,
when entropies ${\mathcal{I}}_q$ or ${\mathcal{S}}_q$ are used, it
is necessary to know them for all $q$ in order to obtain a full
information on a given statistical system~\cite{PJ1}. For
applications in strange attractors the reader may see
Ref.~\cite{HaPr1}, for reconstruction theorems see,
e.g.,~\cite{Re2,PJ1}.

Let us state here some of the basic characteristics of
${\mathcal{D}}_q$. Among properties that are common to both
R\'{e}nyi's and THC entropies we find

\begin{tabbing}
~~~~~(a) ${\mathcal{D}}_q({\mathcal{P}} = \{ 1,0, \ldots,0 \}) =
0$\\
%\\
~~~~~(b) ${\mathcal{D}}_q({\mathcal{P}}) \geq 0$\\
%\\
~~~~~(c) ${\mathcal{D}}_q$ is decisive, i.e.,
${\mathcal{D}}_q(0,1) = {\mathcal{D}}_q(1,0)$\\
%\\
~~~~~(d) ${\mathcal{D}}_q$ is expansible, i.e.,
${\mathcal{D}}_q(p_1, \ldots, p_n) = {\mathcal{D}}_q(0, p_1, \ldots, p_n)$\\
%\\
~~~~~(e) ${\mathcal{D}}_1 = {\mathcal{I}}_1 = {\mathcal{S}}_1 =
{{S}}$\\
%\\
~~~~~(f) ${\mathcal{D}}_q$ involves a single free parameter - $q$\\
%\\
~~~~~(g) ${\mathcal{D}}_q$ is symmetric, i.e.,
${\mathcal{D}}_q(p_1, \ldots, p_n) = {\mathcal{D}}_q(p_{k_{(1)}},
\ldots, p_{k_{(n)}})$\\
%\\
~~~~~(h) ${\mathcal{D}}_q$ is bounded\\[-2mm]
\end{tabbing}

\noindent Among features inherited from R\'{e}nyi's entropy we can
find

\begin{tabbing}
~~~~~(i) ${\mathcal{D}}_q(A) = f^{-1}\left(\sum_k \varrho_k(q)
f({\mathcal{D}}_q(A_k))\right)$\\
%\\
~~~~~(j) ${\mathcal{D}}_q$ is a strictly decreasing function of
$q$, i.e.,
$d{\mathcal{D}}_q/dq \leq 0$, for any $q > 0$\\[-2mm]
\end{tabbing}

\noindent Result (i) follows from the fact that ${\mathcal{D}}_q$
is a monotonically decreasing function of
$\langle\ln{\mathcal{P}}\rangle_q$ and that
$\langle\ln{\mathcal{P}}\rangle_q$ is a monotonically increasing
function of $q$.
% indeed
%
%\begin{eqnarray}
%\frac{d\langle\ln{\mathcal{P}}\rangle_q}{dq} \ = \ \langle
%(\ln({\mathcal{P}}))^2 \rangle_q - \langle \ln({\mathcal{P}})
%\rangle_q^2  \ \geq \ 0 \, . \label{IIIm}
%\end{eqnarray}
%
%The last relation in (\ref{IIIm}) is due to Jensen's inequality.
%Note that $d{\mathcal{D}}_q/dq = 0$ happens only for the
%degenerate case ${\mathcal{P}} = \{1, \ldots, 0\}$.
Finally, properties imprinted from Tsallis entropy include

\begin{tabbing}
~~~~~(k) $\max_{{\mathcal{P}}} {\mathcal{D}}_q({\mathcal{P}}) =
{\mathcal{D}}_q({\mathcal{P}} = \{ 1/n, \ldots, 1/n \}) = \ln_q n$\\
%\\
~~~~~(l) ${\mathcal{D}}_q$ is $q$ non-extensive, i.e.,
${\mathcal{D}}(A\cap B) = {\mathcal{D}}(A) + {\mathcal{D}}(B|A) +
(1-q){\mathcal{D}}(A){\mathcal{D}}(B|A)$\\[-2mm]
\end{tabbing}

The issue of thermodynamic stability will be discussed separately
in Section~\ref{concavity}.
\vspace{-5mm}
%
%%%%%%%%%%%%%%%%%%%%%%%%%%%%%%%%%%%%%%%%%%%%%%%%%%%%%%%%%%%%%
\section{MaxEnt distributions for ${\mathcal{D}}_q$}\label{X}
%%%%%%%%%%%%%%%%%%%%%%%%%%%%%%%%%%%%%%%%%%%%%%%%%%%%%%%%%%%%%
%
\vspace{-4mm}
According to information theory, the MaxEnt principle yields
distributions which reflect least bias and maximum ignorance about
information not provided to a recipient (or observer). Important
feature of the usual Gibbsian MaxEnt formalism is that maximizers
are all grater than zero and that the maximal entropy is a concave
function of the values of the prescribed constraints~\cite{HBC}.

Let us first address the issue of the ${\mathcal{D}}_q$ maximizer.
We start by seeking the conditional extremum of ${\mathcal{D}}_q$
subject to the constraints imposed by the $q$-averaged value of
energy $E$:
%(or any random quantity representing the constant of the
%motion)
%in the form
\\[-7mm]
\begin{eqnarray}
\langle E \rangle_q \ = \ \sum_k \varrho_k(q)E_k\, .
\end{eqnarray}
%
%We will initially keep $r$ not necessary coincident with $q$.
By considering the normalization condition for $p_i$ we should
extremize the functional\\[-7mm]
\begin{eqnarray}
L_{q}({\mathcal{P}}) \ = \ {\mathcal{D}}_q({\mathcal{P}}) - \Omega
\ \frac{\sum_k (p_k)^q E_k}{\sum_k (p_k)^q} - \Phi \sum_k p_k\, ,
\label{Va1}
\end{eqnarray}
with $\Omega$ and $\Phi$ being the Lagrange multipliers. Setting
the derivatives of $L_{q}({\mathcal{P}})$ with respect to $p_1,
\ldots, p_n$ to zero, we obtain\\[-7mm]
\begin{eqnarray}
\frac{\partial L_{q}({\mathcal{P}})}{\partial p_i} \ &=& \ \left.
\left. e^{(q-1)\langle \ln {\mathcal{P}}\rangle_q} \right( q
(\langle \ln {\mathcal{P}}\rangle_q -\ln p_i ) -1 \right)
\frac{(p_i)^{q-1}}{\sum_k (p_k)^q} \nonumber \\
&-& \ \Omega q \left( E_i - \langle E \rangle_q \right)
\frac{(p_i)^{q-1}}{\sum_{k} (p_k)^q} - \Phi \ = \ 0 \, ,
\;\;\;\;\;\;\;\;\;\; i \ = \ 1,2, \ldots , n\, . \label{VIIIa}
\end{eqnarray}
Note that when $q\rightarrow 1$ then (\ref{VIIIa}) reduces to the
usual condition for Shannon's maximizer. This, in turn, ensures
that in the $q \rightarrow 1$ limit the maximizer boils down to
Gibbs's distribution.
%
%%%%%%%%%%%%%%%%%%%%%%%%%%%%%%%%%%%%%%%%%%%%%%%%%%%%%%%%%%%%%%
%\subsection{ The $r=q$ case}
%%%%%%%%%%%%%%%%%%%%%%%%%%%%%%%%%%%%%%%%%%%%%%%%%%%%%%%%%%%%%%
%
To proceed we note that Eq.(\ref{VIIIa}) can be cast to the form\\[-5mm]
\begin{eqnarray}
\Phi  (p_i)^{1-q} \sum_k (p_k)^q \ = \ \left. \left.
e^{(q-1)\langle \ln {\mathcal{P}} \rangle_q }
\right(q\left(\langle \ln {\mathcal{P}} \rangle_q - \ln p_i\right)
-1\right) - q \Omega (E_i - \langle E\rangle_q ) \, .
\label{VIIIb}
\end{eqnarray}
By multiplying both sides by $\varrho_i(q)$, summing over $i$ and
taking the normalization condition $\sum_k p_k =1$, we obtain \\[-8mm]
\begin{eqnarray}
{\mbox{\hspace{-5mm}}}\Phi = -  e^{(q-1)\langle \ln {\mathcal{P}}
\rangle_q } \, \, \Rightarrow \, \, \frac{\ln(-\Phi)}{q-1} \ = \
\langle \ln {\mathcal{P}} \rangle_q \, \, \Rightarrow \, \,
{\mathcal{D}}_q({\mathcal{P}})|_{\mbox{\footnotesize{max}}} \ = \
\frac{1}{q-1}\ ( \Phi + 1) \, . \label{VIIIc}
\end{eqnarray}
Plugging result (\ref{VIIIc}) back into (\ref{VIIIb}) we have\\[-5mm]
\begin{eqnarray}
\sum_k (p_k)^q \ & = & \ (p_i)^{q-1} \left[ q \ln p_i + \left( 1 -
\frac{q \ln(-\Phi)}{q-1} - \frac{q\Omega}{\Phi} \ (E_i - \langle E
\rangle_q) \right) \right] \, , \label{V1}
\end{eqnarray}
which must hold for any index $i$. On the substitution\\[-7mm]
\begin{eqnarray}
{\mathcal{E}}_i \ =  \ 1 - \frac{q \ln(-\Phi)}{q-1} -
\frac{q\Omega}{\Phi} \ \Delta_q E_i\, , \;\;\;\;\;\;\; \Delta_q
E_i \ = \ E_i - \langle E\rangle_q\, ,
\end{eqnarray}
we finally obtain the equation\\[-8mm]
\begin{eqnarray}
\kappa (p_i)^{1-q} \ = \ q \ln p_i \ + \ {\mathcal{E}}_i\, ,
\;\;\;\;\;\;\;\; \sum_k (p_k)^q \equiv \kappa\, . \label{V4}
\end{eqnarray}
This has the solution\\[-7mm]
\begin{eqnarray}
&&p_i \; \, = \;\, \left[  \frac{q}{\kappa (q-1)} \ W\left(
\frac{\kappa (q-1)}{q} \ e^{(q-1)
{\mathcal{E}}_i/q}\right) \right]^{1/(1-q)}\nonumber \\[3mm]
 &&{\mbox{\hspace{4cm}}} = \, \; \exp\left\{ \frac{W\left( \frac{\kappa (q-1)}{q} \
e^{(q-1) {\mathcal{E}}_i/q} \right)}{(q-1)} - {\mathcal{E}}_i/q
\right\}\, , \label{V2}
\end{eqnarray}
with $W(x)$ being the Lambert $W$-function~\cite{CGHJK}.

Some comments are now in order. First, $p_i$'s as prescribed by
(\ref{V2}) are positive for any value of $q > 0$. This is a
straightforward consequence of the following two
identities~\cite{CGHJK}:\\[-5mm]
\begin{eqnarray}
&&W(x) \ = \ \sum_{n=1}^{\infty} \frac{(-1)^{n-1}
n^{n-2}}{(n-1)!}\ x^n \, , \;\;\;\;\;\;\;  W(x) \ = \ x \
e^{-W(x)}\, . \label{V3b}
\end{eqnarray}
Indeed, the first relation ensures that for $x < 0$ also $W(x) <
0$ and hence $W(x)/x > 0$. Thus for $0 < q <1$ the positivity of
$p_i$'s is proven.  Positivity for $q \geq 1$ follows directly
from the second relation. Second, as $q  \rightarrow 1$ the
entropy ${\mathcal{D}}_q \rightarrow {{S}}$ and hence $p_i$'s
defined by (\ref{V2}) approaches the Gibbs distribution. To see
that this is the case,
let us realize that\\[-7mm]
\begin{eqnarray}
\left.\Phi\right|_{q =  1} \ = \  -1\, , \;\;\;
\left.{\mathcal{E}}_i\right|_{q = 1 } \ = \ 1 + {\mathcal{H}} +
\Omega (E_i - \langle E \rangle )\, , \;\;\;\mbox{and} \;\;\;
\left.\kappa \right|_{q = 1}  \ = \ 1\, .
\end{eqnarray}
Then\\[-8mm]
\begin{eqnarray}
{\mbox{\hspace{-6mm}}}\left.p_i\right|_{q = 1}  = \ \exp{\left(1 -
(1 + {\mathcal{H}} + \Omega (E_i - \langle E \rangle)) \right)} \
= \ \exp{\left(\Omega F - \Omega E_i\right)}  \ = \ e^{-\Omega
E_i}/Z \, ,
\end{eqnarray}
which after identification $\left.\Omega\right|_{q = 1} = \beta$
leads to the desired result. Note also that (\ref{V2}) is
invariant under uniform translation of the energy spectrum, i.e.,
the corresponding $p_i$ is insensitive to the choice of the
ground-state energy. Third, there does not seem to be any simple
method for determining $\Phi$ and $\Omega$ in terms of $\langle E
\rangle_q$.
%--- this is true also in the ordinary Gibbs statistics.
In fact, only asymptotic situations for large and vanishingly
small $\Omega$ can be successfully tackled.
%Some progress can be made in the case of
%multifractals~\cite{PJ3} where $\Phi$ and $\Omega$ can be phrased
%in term of experimentally accessible scaling exponents.
For this purpose we briefly remark on the asymptotic behavior of
$p_i$ in regard to $\Omega$.

We first assume that $\Omega \ll 1$
--- ``{\em high-temperature expansion}" --- then from (\ref{V3b}) follows\\[-6mm]
\begin{eqnarray*}
{\mbox{\hspace{-6mm}}}W\!\left( \frac{\kappa(1-q)}{\Phi q}  \ e^{
(q-1)/q}  \exp\left((1-q)\frac{\Omega}{\Phi} \Delta_q E_i\right)
\right) \ \approx \  W\!\left( x \right) \left[ 1 \ - \ (1-q)
\Omega^* \Delta_q E_i \right]\, ,
\end{eqnarray*}
with \\[-9mm]
\begin{eqnarray*}
\Omega^* \ = \ - \frac{\Omega}{ \Phi(W(x)+1) }\, ,\;\;\;\;\;
\;\;\; x \ = \ -\frac{\kappa(q-1)}{\Phi q}
\exp\left(\frac{q-1}{q}\right) \, .
\end{eqnarray*}
The relation (\ref{V2}) then implies that\\
\begin{equation}
{\mbox{\hspace{-5mm}}}p_i  \ = \ \frac{\left[1 \ - \ (1-q)
\Omega^* \Delta_q E_i \right]^{1/(1-q)}}{\sum_k \left[ 1 \ - \
(1-q) \Omega^* \Delta_q E_k \
 \  \right]^{1/(q-1)}} \ = \ Z^{-1} \left[1 \ - \ (1-q) \Omega^*
\Delta_q E_i \right]^{1/(1-q)} , \label{Va2}
\end{equation}
with the partition function\\[-8mm]
\begin{eqnarray}
Z = \sum_k \left[ 1 \ - \ (1-q) \Omega^* \Delta_q E_k  \
\right]^{1/(1-q)} \ = \ \left[ \frac{q}{\kappa (q-1)}
W(x)\right]^{1/(q-1)}\, .
\end{eqnarray}
The distribution (\ref{Va2}) agrees with the so called 3rd version
of thermostatics introduced by Tsallis {\em et al.}~\cite{Ts2}. It
can by also formally identified with the maximizer for
RE~\cite{Ba1}. Clearly, $\Omega^*$ is not a Lagrange multiplier,
but $\Omega^*$ passes to $\beta$  at $q\rightarrow1$ (in fact,
$\Phi \rightarrow -1$, $\Omega \rightarrow \beta $ and $W(x)
\rightarrow 0$ at $q\rightarrow1$). Note also that when $\Omega
=0$ (i.e., no energy constraint) then $p_i =1/n$ which reconfirms
that ${\mathcal{D}}_q$ reaches its maximum for uniform
distribution.

From the physical standpoint it is the asymptotic behavior at
$\Omega|(q-1)/\Phi| \gg 1$
--- ``{\em low-temperature" expansion} --- that is most intriguing. This is because
the branching properties of the Lambert W-function at negative
argument make the structure of ${\mathcal{P}}$ non-trivial. To
this end one can distinguish four distinct situations:\\[-8mm]
\begin{eqnarray*}
a_1)\;\;(q-1) > 0 \;\;\;&\mbox{and}& \;\;\; \Delta_q E_i < 0\, ,
\;\;\;\;\;\;\;  \;\;\;\;\;\;\;
a_2)\;\;(q-1) > 0 \;\;\; \mbox{and} \;\;\; \Delta_q E_i > 0\, , \\
b_1)\;\;(q-1) < 0 \;\;\; &\mbox{and}& \;\;\; \Delta_q E_i < 0\, ,
\;\;\;\;\;\;\;  \;\;\;\;\;\;\;  b_2)\;\;(q-1) < 0 \;\;\;
\mbox{and} \;\;\; \Delta_q E_i > 0\, .
\end{eqnarray*}
Cases $a_1)$ and $a_2)$ are much simpler to start with as the
argument of $W$ is positive. $W$ is then a real and single valued
function which belongs to the principal branch of $W$ known as
$W_0$. When $\Delta_q E_i < 0$ then $a_1)$ implies $W(z)  \approx
z$
and hence \\[-8mm]
\begin{eqnarray}
p_i \ = \ \left(\frac{1}{|\Phi |}\right)^{1/(1-q)}\!e^{-1/q}
\exp\left(-\frac{\Omega}{|\Phi|}\Delta_q E_i\right) \ = \ Z_1^{-1}
\exp\left(-\frac{\Omega}{|\Phi|}\Delta_q E_i\right)\, ,
\label{disa}
\end{eqnarray}
with \\[-9mm]
\begin{eqnarray*}
Z_1\ = \
%\ = \ \sum_{k; \Delta_q E_k < 0}
%\exp\left(-\frac{\Omega}{|\Phi|}\Delta_q E_i\right) \ = \
\left(\frac{1}{|\Phi |}\right)^{1/(q-1)}\!e^{1/q}\, .
\end{eqnarray*}
Note that in this case $p_i$ is of a Boltzmann type. On the other
hand, the $a_2)$ situation implies the asymptotic
expansion~\cite{CGHJK}:\\[-9mm]
\begin{eqnarray}
W(z)\ \approx \ \ln(z) - \ln(\ln(z)) \;\;\; \Rightarrow \;\;\; p_i
\ = \ Z_2^{-1} \left[1 - (1-q)\Omega^* \Delta_q E_i
\right]^{1/(1-q)}\, , \label{disb}
\end{eqnarray}
with \\[-8mm]
\begin{eqnarray*}
{\mbox{\hspace{-6mm}}}Z_2
%\ = \ \sum_{k; \Delta_q E_k > 0}\left[1 - (1-q)\Omega^*
%\Delta_q E_k \right]^{1/(1-q)}
\ = \ \left[\frac{q}{\kappa (q-1)} \ln\!\left( \frac{\kappa
(q-1)}{|\Phi| q} \ e^{ (q-1)/q} \right)\right]^{1/(q-1)}\!\!\! ,
\;\;\; \,\Omega^* \ = \ \frac{\Omega}{|\Phi| \ln\!\left(
\frac{\kappa (q-1)}{|\Phi| q} \exp\left( \frac{q-1}{q} \right)
\right)}\, .
\end{eqnarray*}
Although the distribution (\ref{disb}) formally agrees with
Tsallis {\em et al.} distribution, it cannot be identified with it
as $\Omega^*$ does not tend to $\beta$ in the $q\rightarrow1$
limit. In fact, the limit $q\rightarrow1$ is prohibited as it
violates the ``low-temperature" condition $\Omega|(q-1)/\Phi| \gg
1$. Note particularly that our MaxEnt distribution represents in
the ``low-temperature" regime a heavy tailed distribution with
Boltzmannian outset --- behavior typical, e.g., for income
distributions. When $\Omega$ and $q
> 1$ are fixed one may find $\kappa$ and $\Phi$ from the
normalization condition and sewing condition at $\Delta_q E = 0$.
However, because the ``low-temperature" approximation does not
allow to probe regions with small $\Delta_q E$ one must
numerically optimize the sewing by interpolating the forbidden
parts of $\Delta_q E$ axis~\cite{PJ3}.
%Example of such a numerical simulation
%is presented in Fig.\ref{fig3}
%
%\begin{figure}
%\vspace{4mm} \epsfxsize=8cm
%\centerline{\epsffile{gy6.eps}}
%\vspace{4mm} \caption{A plot of the ``low--temperature" MaxEnt
%distribution (\ref{disa})--(\ref{disb}).  The parameters of the
%plot are chosen in the following way: $\kappa = 0.01$, $\Phi =
%0.68$, $q = 30$ and $\Omega = 0.5$. The distribution is normalized
%to $1$ on the interval ${\Delta}_q E \in [-0.5,0.5]$.}
%\label{fig3}
%\begin{picture}(20,7)
%\put(200,80){ $q$ } \put(80,55){ $p$ } \put(230,155){$\varrho$}
%\end{picture}
%\end{figure}

%To eliminate $\kappa$ from (\ref{V2}) we use the following trick. We
%use the definition $\kappa = \sum_k (p_k)^q$ together with the
%relation (\ref{VIIIc}), then
%
%\begin{eqnarray}
%&&\frac{d\kappa(q)}{dq} \ = \ \sum_k (p_k)^q \ln p_k \ = \ \kappa(q)
%\
%\frac{\ln(-\Phi)}{q-1}\nonumber \\
%&&\frac{d\Phi(q)}{dq} \ = \
%\end{eqnarray}
%

Cases $b_1)$ and $b_2)$ have much richer structure than $a_1)$ and
$a_2)$. This is due to the negativity of the argument that enters
the $W$ function. A remarkable upshot of this is an existence of a
strongly suppressive effect in the occupation of the high-energy
states. In addition, the suppression appears in two different ways
depending on the value of $(1-q)/|\Phi|$. Analogous type of
behavior is know in quantum phase transitions~\cite{Sa1}. Complete
discussion of this phenomenon will be presented in
Ref.~\cite{PJ3}.
\vspace{-5mm}
%
%%%%%%%%%%%%%%%%%%%%%%%%%%%%%%%%%%%%%%%%%%%%%%%%%%%%%%%%%%%%%%
\subsection{Thermodynamic stability --- concavity issue}
\label{concavity}
%%%%%%%%%%%%%%%%%%%%%%%%%%%%%%%%%%%%%%%%%%%%%%%%%%%%%%%%%%%%%%
%
\vspace{-4mm}
In the following we are going to address the issue of
thermodynamic stability. Note that in contrast to
information-theoretic entropy ${\mathcal{D}}_q$,
${\mathcal{D}}_q|_{\mbox{\footnotesize{max}}}$ is the system
entropy, i.e., it depends on the system state variables.
Thermodynamic stability then consists of showing that
${\mathcal{D}}_q|_{\mbox{\footnotesize{max}}}$ is a concave
function of the energy constraint~\cite{HBC}.
%as in the  Gibbsian
%MaxEnt.
So we wish to show that \\[-7mm]
\begin{eqnarray}
\frac{\partial^2
{\mathcal{D}}_q({\mathcal{P}})|_{\mbox{\footnotesize{max}}}}{\partial
\langle E \rangle_q^2} \, = \,
\frac{\partial^2{\mathcal{D}}_q(\langle E \rangle_q)}{\partial
\langle E \rangle_q^2} \, \leq \ 0\, .
%{\mathcal{D}}_q(\lambda \langle E^{(1)} \rangle_q +
%(1-\lambda)\langle E^{(2)} \rangle_q
%)|_{\mbox{\footnotesize{max}}} \ \geq \ \lambda
%{\mathcal{D}}_q(\lambda \langle E^{(1)}
%\rangle_q)|_{\mbox{\footnotesize{max}}} +
%(1-\lambda){\mathcal{D}}_q(\lambda \langle E^{(2)}
%\rangle_q)|_{\mbox{\footnotesize{max}}}\, .
\end{eqnarray}
%.
This can be done by observing that~\cite{CGHJK}\\[-6mm]
\begin{eqnarray}
\frac{d W(x)}{dx} \ = \ \frac{W(x)}{x(W(x)+ 1)} \,\;\;\;\;\;
\mbox{and}\;\;\;\;\; \frac{d^2 W(x)}{dx^2} \ = \ - \frac{W(x)^2
(W(x) + 2)}{x^2(W(x) + 1)^3}\, . \label{24}
\end{eqnarray}
If we combine (\ref{24}) with the fact that
$d{\mathcal{D}}_q({\mathcal{P}})|_{\mbox{\footnotesize{max}}}/d\Phi
= 1/(q-1)$ we
obtain\\[-5mm]
\begin{eqnarray}
\frac{\partial^2
{\mathcal{D}}_q({\mathcal{P}})|_{\mbox{\footnotesize{max}}}}{\partial
\langle E \rangle_q^2} \, = \, (1-q)^2 \frac{|\Phi|}{\Omega}
\langle \ln {\mathcal{P}}\rangle_q \, \leq \, 0\, .
\end{eqnarray}
Thus ${\mathcal{D}}_q$ is thermodynamically stably for any $q$.

\vspace{-7mm}
%
%%%%%%%%%%%%%%%%%%%%%%%%%%%%%%%%%%%%%%%%%%%%
\section{Conclusions}\label{IX}
%%%%%%%%%%%%%%%%%%%%%%%%%%%%%%%%%%%%%%%%%%%%
%
\vspace{-4mm}
In this paper we have reviewed the main aspects of the recently
proposed information measure ${\mathcal{D}}_q$. In contrast to
presently popular generalizations based on deformed entropies, we
have aimed here at a strictly axiomatic approach. This is because
we hold that one cannot proceed the formal generalization of the
entropy in physics by ignoring the consistency with information
theory. As a rule, axiomatic treatments of information measures
have the benefit of a closer passage to operational
characterizations and hence to a systematic use in practical
applications.

We hope that the proposed axiomatics might serve as a novel
playground for $q$-nonextensive systems with embedded
self-similarity. Indeed, our conclusions hint that
${\mathcal{D}}_q$ could play a relevant r\^{o}le in quantum phase
transitions and/or in econophysics.
%This observation is due to  ***.
%Pertinent examples in this context are ***

%The reader may recognize in ${\mathcal{C}}_q$ the generalized
%measure of cross-entropy of Havrda and Charvat~\cite{HaCh} (also
%known as Csisz\'{a}r's measure of directed divergence~\cite{Csi2})
%used in communication theory. For $q=2$ we recover the $\chi^2$
%measure.

%This suggests that the non-extensivity together with
%self-similarity may be important concepts also in information
%theory.

The reader may note that we have not checked ${\mathcal{D}}_q$ for
Lesche's observability criterium~\cite{Leshe} (also known as
experimental robustness). This is because in our view the use of
Lesche's condition as the stability criterion is rather doubtful,
see e.g., Refs.~\cite{PJ25,Yam}. In this connection Yamano's local
stability criterion~\cite{Yam} would seem more appropriate concept
to use. Work along those lines is currently in progress.

Finally we should stress that the presented entropy
${\mathcal{D}}_q$ has many desirable attributes: like THC entropy
it satisfies the nonextensive $q$-additivity, involves a single
parameter $q$, goes over into $S$
%the standard Shannon entropy
in the limit $q \rightarrow 1$, it
complies with thermodynamic stability, continuity, symmetry,
expansivity, decisivity, etc.. On that basis it would appear that
both ${\mathcal{S}}_q$ and ${\mathcal{D}}_q$ have an equal right
to serve as a generalization of statistical thermodynamics.

%%%%%%%%%%%%%%%%%%%%%%%%%%%%%%%%%%%%%%%%%%%%%%%%%%%%%%%%%%%%%%%%%%%%%%%%
%\section*{References}
%%%%%%%%%%%%%%%%%%%%%%%%%%%%%%%%%%%%%%%%%%%%%%%%%%%%%%%%%%%%%%%%%%%%%%%


\vspace{-3mm}
\begin{thebibliography}{999}
%
\vspace{-3mm}
\bibitem{Jaynes57} E.T.~Jaynes, Phys. Rev. {\bf 106} (1957) 171; {\bf
108} (1957) 620.


\bibitem{Fad1} D.K~Faddeyev, Uspekhi Mat. Nauk {\bf 11} (1956)
227 .

\bibitem{SJ1} J.E.~Shore and R.W.~Johnson, IEEE Trans. Inform. Theory {\bf
26} (1980) 26.

\bibitem{Wal1} in, E.T.~Jaynes, {\em Probability Theory, The Logic of
Science} (CUP, Cambridge, 2003).


\bibitem{HaCh} J.H.~Havrda and F.~Charv\'{a}t, Kybernatica
{\bf 3} (1967) 30.


\bibitem{SM1} B.D.~Sharma and D.P.~Mittal, J. Math. Sci. {\bf 10}
(1975) 28.

\bibitem{Re2} A.~R{\'e}nyi,
{\em Selected Papers of Alfred R{\'e}nyi, Vol.2} (Akad{\'e}mia
Kiado, Budapest, 1976).

\bibitem{Kap1} J.N.Kapur, Ind. Jour. Pure and App. Maths. {\bf 17} (1986) 429.


%\bibitem{Urbanik1}K.~Urbanik, Rep. Math. Phys. {\bf 4} (1973)
%289.

\bibitem{Kh1} A.I.~Khinchin, {\em Mathematical Foundations of Information
Theory} (Dover, London, 1957).

\bibitem{Re1} A.~R{\'e}nyi, {\em Probability Theory}
(North-Holland, Amsterdam, 1970).

\bibitem{PJ1} P.~Jizba and T.~Arimitsu, Ann. Phys. {\bf 312}
(2004) 17.


\bibitem{Naudts} see e.g., C.~Tsallis, J. Stat. Phys. {\bf 52} (1988) 479;
Braz. J. Phys. {\bf 29} (1999) 1; J.~Naudts, Physica {\bf A 316}
(2002) 323; G.~Kaniadakis, Phys. Rev. {\bf E 66} (2002) 056125.


\bibitem{Csi1}I.~Csisz{\'a}r, IEEE Trans. Inform. Theory {\bf
41} (1995) 26.

\bibitem{Dar1}Z.~Dar\'{o}czy, Acta Mathematica Academiae Scientiarium
Hungaricae {\bf 15} (1964) 203.

%\bibitem{Ko1} A.~Kolmogorov, Atti della R.  Accademia Nazionale dei Lincei
%{\bf 12} (1930) 388; M.~Nagumo, Japanese Jour. Math. {\bf 7}
%(1930) 71.

%\bibitem{VD2} V.G.~Drinfeld,  Quantum Groups, in: A.~Gleason, ed., {\em Proc.
%Intern. Congr. Math.\/} (Berkeley, 1986) p. 798; M.~Jimbo, Lett.
%Math. Phys. {\bf{10}} (1985) 63.

\bibitem{Oth2} S.~Guias,
{\em Information Theory with Applications\/} (McGraw-Hill, New
York, 1977) .

\bibitem{Ab2} S.~Abe, Phys. Let. {\bf A 271} (2000) 74.


\bibitem{PJ2} P.~Jizba and T.~Arimitsu, Physica
{\bf A 340} (2004) 110.

\bibitem{PJ3} P.~Jizba and T.~Arimitsu, in progress

\bibitem{HaPr1} H.G.E.~Hentschel and I.~Procaccia, Physica {\bf{ D 8}}
(1983) 435.

\bibitem{HBC} H.B.~Callen, {\em Thermodynamics and an introduction to
thermostatistics} (Wiley, New York, 1985).

\bibitem{CGHJK} R.M.~Corless, G.H.~Gonnet, D.E.~Hare, D.J.~Jeffrey
and D.E.~Knuth, Adv. Comput. Math. {\bf 5} (1996) 329.

\bibitem{Ts2}C.~Tsallis, R.S.~Mendes and A.R.~Plastino, Physica {\bf A
261} (1998) 534.

\bibitem{Ba1} A.G.~Bashkirov and A.D.~Sukhanov, JETP {\bf 95} (2002) 440;
A.G.~Bashkirov, Physica {\bf A 340} (2004) 153.

\bibitem{Sa1} S.~Sachdev, {\em Quantum Phase Transitions} (CUP, New York,
1999).

\bibitem{Leshe} B.~Lesche, J. Stat. Phys. {\bf 27} (1982) 419.

\bibitem{PJ25} P.~Jizba and T.~Arimitsu, Phys. Rev. {\bf E 69}
(2004) 026128.

\bibitem{Yam} T.~Yamano, Phys. Lett. {\bf A 329} (2004) 268.






%\bibitem{BS1}Ch.~Beck and F.~Schlogl, {\em Thermodynamics of Chaotic
%Systems, An Introduction} (Cambridge University Press, Cambridge,
%1993).

\end{thebibliography}
\end{document}